\documentclass[a4paper,11pt]{article}
\usepackage{pos}

\title{EFT description of the muon magnetic dipole moment}

\author*[1]{Jason~Aebischer}

\emailAdd{jason.aebischer@physik.uzh.ch}

\affiliation[1]{Physik-Institut, Universit\"at Z\"urich, CH-8057 Z\"urich, Switzerland}

\abstract{An Effective Field Theory (EFT) analysis of the magnetic moment of the muon is discussed. The expression for the dipole moment is given in terms of operator coefficients of the low-energy effective field theory (LEFT) and the Standard Model effective field theory (SMEFT), where one-loop renormalization group improved perturbation theory, the one-loop matching from SMEFT onto LEFT as well as one-loop lepton matrix elements of the effective-theory operators were taken into account. Interestingly only a very limited set of the SMEFT operators is able to explain the current deviation of the magnetic moment of the muon from its Standard Model expectation.}

\FullConference{%
  *** The European Physical Society Conference on High Energy Physics (EPS-HEP2021), ***\\
  *** 26-30 July 2021 ***\\
  *** Online conference, jointly organized by Universität Hamburg and the research center DESY ***
}

\def\tcwc#1#2#3{\mathcal{\widetilde C}_{\substack{#1 \\ #3}}^{#2}}
\def\tlwc#1#2#3#4{\widetilde  L_{\substack{#1 \\ #4}}^{#3,#2}}
\renewcommand{\Re}{\mathrm{Re}}

\newcommand{\nn}{\nonumber\\}
\newcommand{\GeV}{\;\mathrm{GeV}}
\newcommand{\TeV}{\;\mathrm{TeV}}
\DeclareUnicodeCharacter{2212}{\textendash}

\begin{document}
\maketitle

\section{Introduction}

The longstanding anomaly in the anomalous magnetic moment of the muon, $a_\mu$, still persists after the new measurement of the Muon g-2 collaboration at Fermilab this year \cite{Muong-2:2021vma}. The experimental value, when averaged with the result from the E821 experiment at Brookhaven \cite{Bennett:2006fi}, is given by
\begin{align}
	a_\mu^\mathrm{exp} &= 116\,592\,061(41) \times 10^{-11},
\end{align}
compared to the Standard Model (SM) prediction of~\cite{Aoyama:2020ynm}
\begin{align}
	a_\mu^\mathrm{SM} &= 116\,591\,810(43) \times 10^{-11}\,.
\end{align}
The difference between the experimental value and the SM prediction is given by
\begin{align}
	\label{eq:MuonDiscrepancy}
	\Delta a_\mu &= a_\mu^\mathrm{exp} - a_\mu^\mathrm{SM} = 251(59) \times 10^{-11}\,,
\end{align}
which corresponds to a tension of $4.2\,\sigma$. Many explanations were proposed to resolve this anomaly by assuming specific models such as Leptoquark-, $Z^\prime$- or suspersymmetric models. Here we discuss a model-independent approach to describe the tension between theory and experiment by employing Effective Field Theories (EFTs). Instead of assuming new fields and their couplings to the SM particles we compute the anomalous magnetic moment of the muon in the context of the most general EFTs involving the SM fields. More precisely, the observable is expressed in terms of the Wilson coefficients of the underlying EFT. Such a {\it master formula} allows to compute the observable for any given New Physics (NP) model, after matching the model onto the underlying EFT. To derive the master formulae for the anomalous magnetic moment above as well as below the electroweak (EW) scale we assumed the most general EFTs of the SM.

Above the EW scale the corresponding EFT is the Standard Model Effective Theory (SMEFT) \cite{Grzadkowski:2010es}.\footnote{If the Higgs field is assumed to be a singlet under $SU(2)_L$ the SMEFT can be generalized to the Higgs Effective Field Theory (HEFT).} The SMEFT includes all operators of mass dimension $\leq 6$ that contain SM fields and which are invariant under the SM gauge group $SU(3)_{\text{c}}\times SU(2)_{\text{L}}\times U(1)_{\text{Y}}$. The complete Renormalization Group (RG) running for all the SMEFT parameters is known at one-loop~\cite{Jenkins:2013zja,Jenkins:2013wua,Alonso:2013hga,Alonso:2014zka}.

After spontaneous symmetry breaking, the SM gauge symmetry is broken into the smaller gauge group $SU(3)_{\text{c}}\times U(1)_{\text{em}}$, including only QCD and QED as gauge interactions. The corresponding effective field theory valid below the EW scale is the low-energy effective field theory (LEFT). The full operator basis up to dimension six of the LEFT is known~\cite{Jenkins:2017jig} as well as the complete QCD and QED running of all the Wilson coefficients~\cite{Jenkins:2017dyc,Aebischer:2017gaw}.

Finally, the complete tree-level and one-loop matching from the SMEFT onto the LEFT has been computed~\cite{Dekens:2019ept,Aebischer:2015fzz}. All these results were taken into account to obtain the most general master formulae for magnetic as well as electric dipole moments for the muon and the electron in Ref.~\cite{Aebischer:2021uvt}. A previous EFT result involving the muon $(g-2)$ in the SMEFT can be found in \cite{Crivellin:2013hpa}. A recent analysis studying in addition neutron electric dipole operators in the SMEFT is given in \cite{Kley:2021yhn} and an EFT analysis in the $\nu$SMEFT can be found in \cite{Cirigliano:2021peb}. Here we describe the results obtained in Ref.~\cite{Aebischer:2021uvt} but limit the discussion to the anomalous magnetic moment of the muon. In the spirit of Ref.~\cite{Aebischer:2018quc,Aebischer:2018csl,Aebischer:2020dsw,Aebischer:2021raf,Aebischer:2021hws} we present the master formulae for $\Delta a_\mu$ in terms of the LEFT coefficients in Sec.~\ref{sec:LEFT} and for the SMEFT Wilson coefficients in Sec.~\ref{sec:SMEFT}. We summarize the obtained results in Sec.~\ref{sec:summary}.

\section{LEFT master formula}\label{sec:LEFT}

As discussed in detail in Ref.~\cite{Aebischer:2021uvt} in a first step we compute the anomalous magnetic moment at the low scale  $\mu=2\GeV$ at one-loop in the LEFT. The corresponding master formula for $\Delta a_\mu$ is reported in \cite{Aebischer:2021uvt}. In a next step we consider the one-loop RG running from the low scale up to $\mu=60\GeV$ in the LEFT. We choose this scale to be in between the low and the EW scale, such that QCD and QED RGE effects in the LEFT can be studied. The running of the Wilson coefficients is performed using the Python package \texttt{wilson} \cite{Aebischer:2018bkb} and we adopt the \texttt{WCxf} \cite{Aebischer:2017ugx} standard for the Wilson coefficients in the JMS basis. The resulting master formula for $\Delta a_{\mu}$ at the scale of $\mu=60\GeV$ takes the form
\begin{align}\label{eq:LEFT}
& \Delta a_{\mu}^{60\GeV} =\Re\Bigg[
{2.2} \times 10^{-2} {\widetilde L}_{\substack{e\gamma\\ \mu \mu}}
 - {5.3} \times 10^{-5}  \tlwc{ed}{RR}{T}{\mu\mu bb}   \nn
& + \left({3.5} +0.65 c_T^{(c)} \right) \times 10^{-5} \tlwc{eu}{RR}{T}{\mu\mu cc  }
 +{ 9.0} \times 10^{-6}  \tlwc{ee}{RR}{S}{\mu \tau \tau \mu}
- 1.4 \times 10^{-6}  \tlwc{ee}{LR}{V}{\mu \tau \tau \mu}  \nn
& + { 9.8} \times 10^{-7}  \tlwc{ee}{RR}{S}{\mu \mu\mu \mu}
 - \left(10 c_T - {0.64} \right) \times 10^{-7}   \tlwc{eu}{RR}{T}{\mu\mu uu } \nn
& + \left(5.0  c_T - {14} \right) \times 10^{-7}  \tlwc{ed}{RR}{T}{\mu\mu ss }
 + \left(5.0 c_T - {0.70}\right) \times 10^{-7}  \tlwc{ed}{RR}{T}{\mu\mu dd } \nn
& - { 1.6} \times 10^{-7}  \tlwc{ee}{RR}{S}{\mu\mu \tau \tau }
 - \left({5.9} +2.3 c_T^{(c)}+0.45 c_S^{(c)} \right) \times 10^{-8} \tlwc{eu}{RR}{S}{\mu \mu cc } \nn
& - {8.0} \times 10^{-8} \tlwc{ee}{LR}{V}{\mu \mu \mu \mu }
 - {3.3} \times 10^{-8}  \tlwc{ed}{RR}{S}{\mu\mu b b }
 -2.4  \times 10^{-8} \tlwc{ee}{RR}{S}{\mu\mu \mu \mu }
 + {8.8} \times 10^{-9}  \tlwc{ee}{RR}{S}{\mu e e \mu }\nn
& - 4.5 \times 10^{-9} \widetilde c_S^{(c)} \tlwc{eu}{RL}{S}{\mu \mu cc }
+ 3.5 \times 10^{-9} c_T  \tlwc{eu}{RR}{S}{\mu \mu uu }
 - {1.2} \times 10^{-9}  \tlwc{ed}{RR}{S}{\mu \mu ss  }
\Bigg]\,,
\end{align}
with the $\mathcal{O}(1)$ low-energy constants $c_T,\,c_T^{(c)}\,,c_S^{(c)}\,,\widetilde c_S^{(c)}$ defined in \cite{Aebischer:2021uvt} and where we have used the complex and dimensionless Wilson coefficients ${\widetilde L}_i$, defined by
\begin{equation}
 {\widetilde L}_i\equiv \Lambda^{-d_i}{L}_i(\mu=\Lambda)\,,
\end{equation}
with the operator dimension $d_i$ and the NP scale $\Lambda$.
 Furthermore, in eq.~\eqref{eq:LEFT} and in the following we only retain contributions of at least $10^{-9}$ since the corresponding Wilson coefficients are assumed to be $\sim \mathcal{O}(1)$.

\begin{figure}[tbp]
	\centering
\includegraphics[scale=0.6]{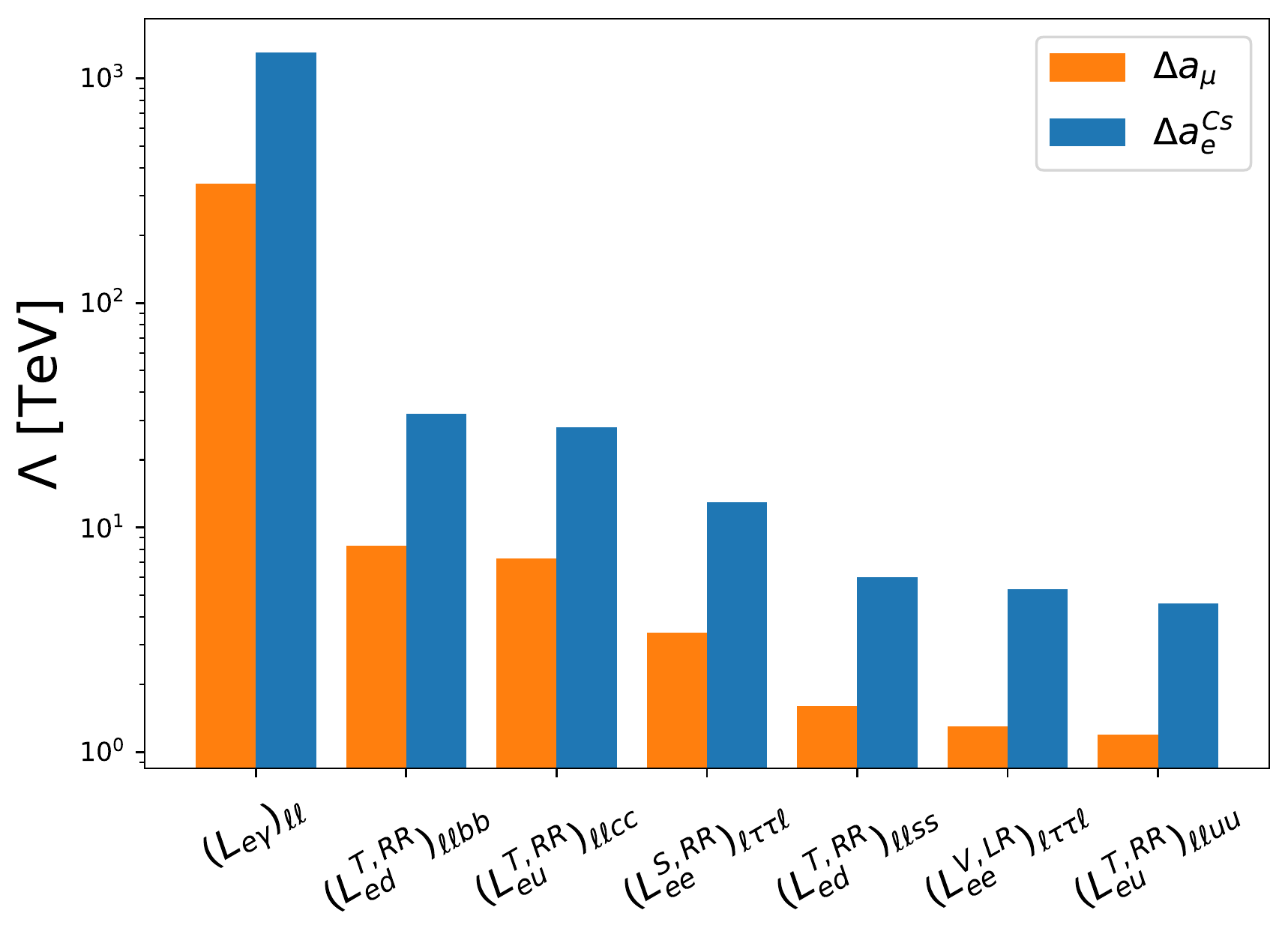}
\caption{The largest scales probed from the Wilson coefficients in eq.~\eqref{eq:LEFT} when imposing the anomalous magnetic moments of the muon (orange) and of the electron (blue).
}\label{fig:barplot}
\end{figure}

 The most important contribution to $a_\mu$ comes from the dipole operator represented by ${\widetilde L}_{\substack{e\gamma\\ \mu \mu}}$, followed by different tensor, scalar and vector four-fermi operators. In order to obtain some intuition about how large the NP scale might be, one can consider individual Wilson coefficients in eq.~\eqref{eq:LEFT}, while setting all the others to zero. Including the LEFT running one can then find the NP scale $\Lambda$ at which the corresponding Wilson coefficient needs to be generated, to give a large enough contribution to $\Delta a_\mu$. In Fig.~\ref{fig:barplot} we show the first seven Wilson coefficients of eq.~\eqref{eq:LEFT} which have the largest impact on the anomalous magnetic moment, together with the NP scale (orange bars) that they probe. The size of the various NP scales ranges from $\sim 100\TeV$ for the dipole operator to several$\TeV$ for the different four-fermi operators. The blue bars show the corresponding constraints when the anomalous magnetic moment of the electron is considered.

\section{SMEFT master formula}\label{sec:SMEFT}

In the SMEFT above the EW scale the picture simplifies drastically. Taking the running from $60\GeV$ to the EW scale, the tree-level and one-loop matching, as well as the complete one-loop SMEFT running into account the master formula for the anomalous magnetic moment of the muon at $10\TeV$ in terms of SMEFT Wilson coefficients reads
\begin{align}
\Delta a_{\mu}^{10\TeV} =\Re\Bigg[ & 1.7 \times 10^{-6}\tcwc{eB}{}{\mu\mu}-{9.2} \times 10^{-7}\tcwc{eW}{}{\mu\mu}- {2.2} \times 10^{-7}\tcwc{\ell equ}{(3)}{\mu\mu33} \nn
  & - \left( {2.5} + 0.22 c_T^{(c)} \right) \times 10^{-9}\tcwc{\ell equ}{(3)}{\mu\mu22}
\Bigg]\,.
\end{align}
Interestingly, only a handful of parameters at the high scale give a large enough contribution to explain the current $(g-2)_\mu$ anomaly: Besides the two SMEFT dipole operators also the tensor operators containing top quarks as well as charm quarks give a sizable contribution to $a_{\mu}$. The semileptonic operators can for example be generated at the high scale in various Leptoquark models as discussed in \cite{Aebischer:2021uvt}.

\section{Summary}\label{sec:summary}
We have discussed general master formulae for the anomalous magnetic dipole moment of the muon in the LEFT below the EW scale and in the SMEFT above the EW scale. In the derivation we included the one-loop running below the EW scale, the one-loop matching at the EW scale as well as the full one-loop running in the SMEFT. Whereas at $\mu=60\GeV$ many different LEFT operators give sizable contributions to $a_\mu$, at $\mu=10\TeV$ only two dipole and two semi-leptonic out of the $\sim 2500$ SMEFT operators have a large enough impact to resolve the current $(g-2)_\mu$ anomaly.

\section*{Acknowledgments}
I thank Wouter Dekens and Peter Stoffer for useful discussions and comments on the manuscript.

\small
\bibliographystyle{unsrt}
\bibliography{Literature}

\begin{thebibliography}{10}

\bibitem{Muong-2:2021vma}
T.~Albahri et~al.
\newblock {Measurement of the anomalous precession frequency of the muon in the
  Fermilab Muon $g−2$ Experiment}.
\newblock {\em Phys. Rev. D}, 103(7):072002, 2021.

\bibitem{Bennett:2006fi}
G.W. Bennett et~al.
\newblock {Final Report of the Muon E821 Anomalous Magnetic Moment Measurement
  at BNL}.
\newblock {\em Phys. Rev. D}, 73:072003, 2006.

\bibitem{Aoyama:2020ynm}
T.~Aoyama et~al.
\newblock {The anomalous magnetic moment of the muon in the Standard Model}.
\newblock {\em Phys. Rept.}, 887:1--166, 2020.

\bibitem{Grzadkowski:2010es}
B.~Grzadkowski, M.~Iskrzy\'nski, M.~Misiak, and J.~Rosiek.
\newblock {Dimension-Six Terms in the Standard Model Lagrangian}.
\newblock {\em JHEP}, 10:085, 2010.

\bibitem{Jenkins:2013zja}
Elizabeth~E. Jenkins, Aneesh~V. Manohar, and Michael Trott.
\newblock {Renormalization Group Evolution of the Standard Model Dimension Six
  Operators I: Formalism and lambda Dependence}.
\newblock {\em JHEP}, 10:087, 2013.

\bibitem{Jenkins:2013wua}
Elizabeth~E. Jenkins, Aneesh~V. Manohar, and Michael Trott.
\newblock {Renormalization Group Evolution of the Standard Model Dimension Six
  Operators II: Yukawa Dependence}.
\newblock {\em JHEP}, 01:035, 2014.

\bibitem{Alonso:2013hga}
Rodrigo Alonso, Elizabeth~E. Jenkins, Aneesh~V. Manohar, and Michael Trott.
\newblock {Renormalization Group Evolution of the Standard Model Dimension Six
  Operators III: Gauge Coupling Dependence and Phenomenology}.
\newblock {\em JHEP}, 04:159, 2014.

\bibitem{Alonso:2014zka}
Rodrigo Alonso, Hsi-Ming Chang, Elizabeth~E. Jenkins, Aneesh~V. Manohar, and
  Brian Shotwell.
\newblock {Renormalization group evolution of dimension-six baryon number
  violating operators}.
\newblock {\em Phys. Lett. B}, 734:302--307, 2014.

\bibitem{Jenkins:2017jig}
Elizabeth~E. Jenkins, Aneesh~V. Manohar, and Peter Stoffer.
\newblock {Low-Energy Effective Field Theory below the Electroweak Scale:
  Operators and Matching}.
\newblock {\em JHEP}, 03:016, 2018.

\bibitem{Jenkins:2017dyc}
Elizabeth~E. Jenkins, Aneesh~V. Manohar, and Peter Stoffer.
\newblock {Low-Energy Effective Field Theory below the Electroweak Scale:
  Anomalous Dimensions}.
\newblock {\em JHEP}, 01:084, 2018.

\bibitem{Aebischer:2017gaw}
Jason Aebischer, Matteo Fael, Christoph Greub, and Javier Virto.
\newblock {B physics Beyond the Standard Model at One Loop: Complete
  Renormalization Group Evolution below the Electroweak Scale}.
\newblock {\em JHEP}, 09:158, 2017.

\bibitem{Dekens:2019ept}
Wouter Dekens and Peter Stoffer.
\newblock {Low-energy effective field theory below the electroweak scale:
  matching at one loop}.
\newblock {\em JHEP}, 10:197, 2019.

\bibitem{Aebischer:2015fzz}
Jason Aebischer, Andreas Crivellin, Matteo Fael, and Christoph Greub.
\newblock {Matching of gauge invariant dimension-six operators for $b\to s$ and
  $b\to c$ transitions}.
\newblock {\em JHEP}, 05:037, 2016.

\bibitem{Aebischer:2021uvt}
Jason Aebischer, Wouter Dekens, Elizabeth~E. Jenkins, Aneesh~V. Manohar, Dipan
  Sengupta, and Peter Stoffer.
\newblock {Effective field theory interpretation of lepton magnetic and
  electric dipole moments}.
\newblock 2 2021.

\bibitem{Crivellin:2013hpa}
Andreas Crivellin, Saereh Najjari, and Janusz Rosiek.
\newblock {Lepton Flavor Violation in the Standard Model with general
  Dimension-Six Operators}.
\newblock {\em JHEP}, 04:167, 2014.

\bibitem{Kley:2021yhn}
Jonathan Kley, Tobias Theil, Elena Venturini, and Andreas Weiler.
\newblock {Electric dipole moments at one-loop in the dimension-6 SMEFT}.
\newblock 9 2021.

\bibitem{Cirigliano:2021peb}
Vincenzo Cirigliano, Wouter Dekens, Jordy de~Vries, Kaori Fuyuto, Emanuele
  Mereghetti, and Richard Ruiz.
\newblock {Leptonic anomalous magnetic moments in \ensuremath{\nu} SMEFT}.
\newblock {\em JHEP}, 08:103, 2021.

\bibitem{Aebischer:2018quc}
Jason Aebischer, Christoph Bobeth, Andrzej~J. Buras, Jean-Marc G\'erard, and
  David~M. Straub.
\newblock {Master formula for $\varepsilon'/\varepsilon$ beyond the Standard
  Model}.
\newblock {\em Phys. Lett. B}, 792:465--469, 2019.

\bibitem{Aebischer:2018csl}
Jason Aebischer, Christoph Bobeth, Andrzej~J. Buras, and David~M. Straub.
\newblock {Anatomy of $\varepsilon '/\varepsilon $ beyond the standard model}.
\newblock {\em Eur. Phys. J. C}, 79(3):219, 2019.

\bibitem{Aebischer:2020dsw}
Jason Aebischer, Christoph Bobeth, Andrzej~J. Buras, and Jacky Kumar.
\newblock {SMEFT ATLAS of $\Delta$F = 2 transitions}.
\newblock {\em JHEP}, 12:187, 2020.

\bibitem{Aebischer:2021raf}
Jason Aebischer, Christoph Bobeth, Andrzej~J. Buras, Jacky Kumar, and
  Miko\l{}aj Misiak.
\newblock {General non-leptonic $\Delta F=1$ WET at the NLO in QCD}.
\newblock 7 2021.

\bibitem{Aebischer:2021hws}
Jason Aebischer, Christoph Bobeth, Andrzej~J. Buras, and Jacky Kumar.
\newblock {BSM Master Formula for $\varepsilon'/\varepsilon$ in the WET Basis
  at NLO in QCD}.
\newblock 7 2021.

\bibitem{Aebischer:2018bkb}
Jason Aebischer, Jacky Kumar, and David~M. Straub.
\newblock {Wilson: a Python package for the running and matching of Wilson
  coefficients above and below the electroweak scale}.
\newblock {\em Eur. Phys. J. C}, 78(12):1026, 2018.

\bibitem{Aebischer:2017ugx}
Jason Aebischer et~al.
\newblock {WCxf: an exchange format for Wilson coefficients beyond the Standard
  Model}.
\newblock {\em Comput. Phys. Commun.}, 232:71--83, 2018.

\end{thebibliography}

\end{document}